# On the Potential of Fourier-Encoded Saturation Transfers for Sensitizing Solid-State Magic-Angle Spinning NMR Experiments

Michael J. Jaroszewicz, Mihajlo Novakovic, and Lucio Frydman*

Department of Chemical and Biological Physics, Weizmann Institute of Science, Rehovot, Israel, 7610001.

* Email: lucio.frydman@weizmann.ac.il

**ABSTRACT**. Chemical exchange saturation transfer (CEST) is widely used for enhancing the solution NMR signatures of magnetically-dilute spin pools; in particular species at low concentrations undergoing chemical exchanges with an abundant spin pool. CEST's main feature involves encoding and then detecting the weak NMR signals of the magnetically dilute spin pools on a magnetically abundant spin pool of much easier detection – for instance the protons of $H_2O$. Inspired by this method, we propose and exemplify a methodology to enhance the sensitivity of magic-angle spinning (MAS) solid-state NMR spectra. Our proposal uses the abundant $^1H$ reservoir arising in organic solids as the magnetically abundant spin pool, and relies on proton spin diffusion *in lieu* of chemical exchange to mediate polarization transfer between a magnetically dilute spin pool and this magnetically abundant spin reporter. As an initial test of this idea we target the spectroscopy of naturally-abundant $^{13}C$, and rely on a Fourier-encoded version of the CEST experiment for achieving broadbandness in coordination with both MAS and heteronuclear decoupling – features normally absent in CEST. Arbitrary evolutions of multiple $^{13}C$ sites can thus be imprinted on the entire $^1H$ reservoir, which is subsequently detected. Theoretical predictions suggest that orders-of-magnitude signal enhancements should be achievable in this manner – on the order of the ratio between the $^{13}C$ and the $^1H$ reservoirs' abundances. Experiments carried out under magic-angle-spinning conditions evidenced *ca*. 5-10× enhancements. Further opportunities and challenges arising in this *Fourier-Encoded Saturation Transfer* (FEST) MAS NMR approach are briefly discussed.

**TOC GRAPHICS**

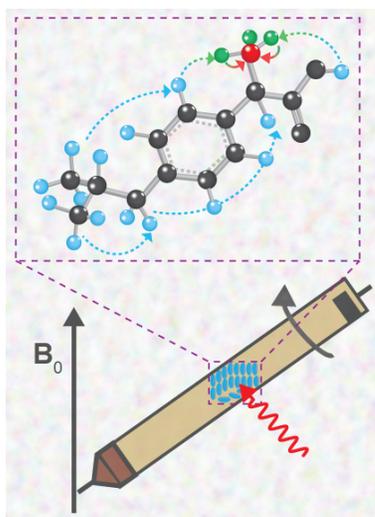

Sensitivity enhancement of dilute $X$ heteronuclei: Here, the spin-polarization of strongly $X$-coupled $^1$Hs (green) is first depleted in an $X$ shift-dependent manner, which is then relayed to the extended and largely $X$-decoupled $^1$H reservoir (blue) by $^1$H-$^1$H spin diffusion. Looping these two processes multiples times and then detecting the resulting $^1$H depletion results in signal enhancement.

**Introduction**

Chemical exchange saturation transfer (CEST) is a widely used technique for enhancing the sensitivity of liquid-state NMR spectra;[1–7] its main applications involve the magnified detection of metabolites and bio-macromolecules *in vivo*,[8–11,24-25] the detection of "invisible" states in high-resolution biomolecular NMR,[12–16] studies of exchange and enzymatic phenomena *in vitro*,[17–20] and NMR enhancements of hyperpolarized substrates.[21–23] CEST involves selectively saturating weak NMR resonances of labile or otherwise interconverting sites, and then relying on chemical exchange processes to transfer this site-selective saturation onto a much stronger NMR resonance. By leveraging the fact that the rate $k_{exch}$ of this exchange process can be much faster than the longitudinal relaxation rate $1/T_1$ of the receiving pool, which is ultimately excited and detected, this saturation-transfer principle can lead to very large sensitivity enhancements on the order of $k_{exch} \cdot T_1$. It follows that, when discussing CEST experiments, it is useful to classify the chemical exchanging spin system into two distinct pools based on their relative populations and on their overall NMR receptivity. Spins giving NMR signals that are of interest, but are weak as a result from low natural abundance, small gyromagnetic ratios, low concentrations, and/or combinations of all these, constitute what we will call the *magnetically dilute* spin pool. This will in turn exchange information with the *magnetically abundant* spin pool, a highly populated and easier to detect reservoir (like the protons in water). Given the ease and robustness with which CEST can be implemented, as well as the large gains in sensitivity that it can provide, CEST has since been exploited in the above-mentioned variety of liquid-state and *in vivo* NMR scenarios –providing a wealth of information pertaining to structure and dynamics.

As originally introduced, CEST is a continuous-wave (CW), frequency-domain experiment targeting what is usually a peak of interest whose resonance frequency is *a priori* known.

*Frequency Labeled EXchange* (FLEX)[22,26] endows CEST with the broadbandness associated with time-domain experiments. In FLEX, the resonances of the dilute pool are not saturated, but rather amplitude-modulated by a pair of selective excitation/storage pulses that avoid perturbing the abundant spin reservoir. These pulses are separated by a $t_1$ delay that is incremented in a normal 2D NMR fashion[27] but which, unlike conventional 2D NMR, is looped multiple times before the final observation. This allows the chemical exchange process to transfer the $t_1$ information onto the abundant reservoir as a magnified amplitude modulation that grows with the number of loops; applying an observation pulse on the abundant reservoir then enables, after Fourier Transform (FT) *vs.* $t_1$, the detection of the dilute pool NMR spectrum with an increased, CEST-like sensitivity. By departing from the original CW saturation scheme, FLEX provides this experiment with additional flexibility. Recently, for example, sensitivity-enhanced solution-state NMR pulse sequences have been developed that exploit this approach for increasing the SNR of not only labile protons, but also the SNR of non-labile (*e.g.*, aliphatic) protons[28,29] as well as non-labile heteronuclei ($^{13}$C, $^{15}$N)[28] for a variety of systems including carbohydrates, amino acids, and intrinsically disordered proteins. Some of these experiments also depart from traditional CW CEST approaches in that following the $t_1$ time-domain encoding, they rely not only on chemical exchange but also on coherent (*e.g.*, TOCSY, INEPT) polarization transfer segments to achieve their substantial signal enhancements in either homonuclear or heteronuclear systems, starting from either labile or non-labile spins.[30] Inspired by these solution-state RElayed-FLEX (REFLEX) sequences,[28-30] we herein explore and test a protocol for implementing conceptually similar experiments, but aimed at acquiring NMR spectra from solid samples undergoing magic-angle-spinning (MAS) and heteronuclear decoupling. As in the solution-state REFLEX counterparts, the aim of these methods will be to enhance 2D heteronuclear spin correlation (HETCOR) experiments involving dilute nuclei like $^{13}$C; unlike the solution-state NMR cases, no chemical exchanges with a dominating solvent will be available for use, and acquisitions will have to be done under the stringent decoupling and spinning manipulations that are needed for collecting high-resolution NMR spectra from powdered solids.

**Fourier-Encoded Saturation Transfer in High-Resolution Solid-State NMR**

*A Fourier-encoded, solid-state NMR version of FLEX.* To extend the solution-phase saturation transfer principles described above to solids undergoing both heteronuclear $^1$H decoupling and MAS, we put forward the *Fourier-Encoded Saturation Transfer* (FEST) experiment. FEST is in principle applicable to the observation of any *X*-nucleus surrounded by a $^1$H reservoir; here we target dilute $^{13}$C surrounded by an ensemble of abundant protons as a representative example. The are several conceptual elements of FEST experiments that have both similarities and differences to the elements involved in solution-state CEST/FLEX NMR; hence they are introduced in this paragraph one by one. The enabling component of FEST are dipolar couplings –both those between the dilute $^{13}$Cs and its neighboring, strongly coupled $^1$Hs, as well as those between these $^{13}$C-coupled $^1$Hs and the $^1$H bulk ensemble at large. $^{13}$C-$^1$H dipolar couplings are normally used in conventional 2D HETCOR solids NMR for sensitizing experiments, for instance upon pre-polarizing the $^{13}$C *via* cross-polarization (CP), and/or when transferring back the $^{13}$C encoding to perform $^1$H-detection with enhanced sensitivity.[31–36] However, by virtue of their separation and of the low $^{13}$C natural abundance, these dipolar-driven processes will involve only a small fraction of the total protons in the sample, as most $^1$Hs are not coupled strongly enough to the carbon in order to participate in polarization transfers. Therefore, this abundant $^1$H spin polarization goes largely unused. FEST seeks to exploit this large portion of the abundant $^1$H spin system as part of the signal enhancement process, by making *it* the reporter of the $^{13}$C NMR time-domain FID. To do so the proposed experiment relies on the REFLEX-inspired approach introduced in **Scheme 1**. This depicts a typical organic solid in terms of three spin pools, each having varying populations and receptivities schematically illustrated by the size of their circles: the $^{13}$C, the $^1$H(s) that is(are) strongly coupled to the $^{13}$C, and a more abundant, essentially $^{13}$C-decoupled $^1$H reservoir, that will make the bulk of the NMR signal in a $^1$H-detected experiment. Further, it is assumed that the distinct dipolar couplings that exist between these various spin pools allows each to be individually addressed with suitable radio-frequency (RF) manipulations; for instance, CP can be used to transfer polarization between $^{13}$C and its directly dipole-coupled $^1$Hs, while spontaneous or RF-driven spin diffusion enables the communication between the latter $^1$Hs and the much larger portion of the abundant $^1$H spin pool (shown in blue).

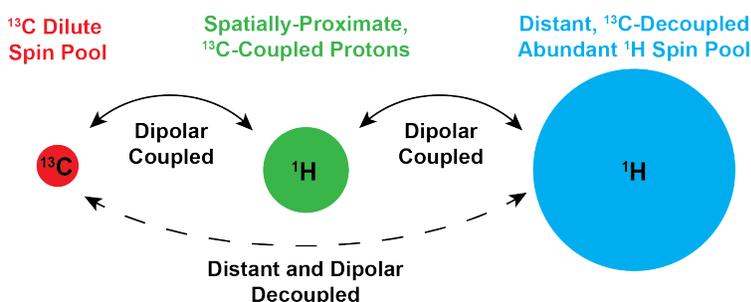

**Scheme 1**. Schematic representation of how a typical organic solid can be formalized in terms of three distinct spin pools (indicated by the different colors), each with varying populations and receptivities (indicated by the size of the circles). The system entails a $^{13}$C (red) that is dipole-coupled to a proximate proton (green), with the latter dipolar-coupled with an abundant $^1$H spin pool (in blue) that is not directly coupled to the heteronucleus, but whose observation is key for enhancing the NMR signal of the $^{13}$C spin.

With this as background, the procedure by which we propose to use the latter for efficiently enhancing the dilute $^{13}$C NMR signature is described in **Figures 1a-1d**. According to this scheme, (1) the proximate $^1$H pool is first used to polarize the $^{13}$C; (2) the latter's time-domain evolution is triggered, encoded over a time $t_1$, and then passed back onto $^1$H$^{(proximate)}$ as an amplitude modulation; (3) a $^1$H-$^1$H spin-diffusion interval is allowed to proceed whereby this proximate $^1$H pool depolarizes the distant $^1$H$^{(abundant)}$ to a degree that reflects the $^{13}$C $t_1$ evolution while getting repolarized in exchange; (4) the whole process is repeated several times, so as to imprint the $^{13}$C evolution onto the bulk $^1$H reservoir to the maximum possible extent. Step (2) acts here as the Fourier-encoding module in FLEX, while steps (3) and (4) would act as analogues of the chemical exchange saturation transfer process amplifying the signal. The $^{13}$C signal magnification that these processes bring about will depend on the number of times that the overall process is repeated, on the degree of $^{13}$C *vs.* $^1$H (bulk) dilution, on the strength of the couplings between the various pools, on the longitudinal $^1$H and $^{13}$C $T_1$ and $T_{1\rho}$ relaxation-time constants, and on the efficiency with which all these processes can be implemented while under the high-resolution requirements of solid-state $^{13}$C NMR.

**Figure 1e** introduces a time-domain pulse sequence that could be used for magnifying in this manner the chemical shift modulation of a dilute $^{13}$C spin pool on a more abundant bulk $^1$H reservoir –while remaining compatible with the decoupling and MAS demands imposed by high resolution $^{13}$C observations. The steps that are involved in the ensuing indirect-detection sequence, as well as their approximate correlation with the processes introduced in **Figure 1a-1d**, include: (1) An initial block where polarization is received by the $^{13}$C from its proximate, dipolar-coupled $^1$H *via* an optimized CP process, which is long enough to be effective, but also short so as to not incur detrimental $T_{1\rho}$ $^1$H relaxation losses. Particularly important is the preservation of the spin polarization belonging to the abundant proton pool, which is therefore stored post-CP along the +z axis. (2) This is followed by a mixing interval $\tau_{mix}$ that is sufficiently long for enabling $^1$H-$^1$H spin-diffusion yet short *vs* the $^{13}$C $T_1$, so that the $^{13}$C remain fully polarized at the end of $\tau_{mix}$, while spin polarization from $^1$H$^{(abundant)}$ spontaneously repolarizes the depleted $^{13}$C-coupled proton through spin diffusion.[37] (3) In a process that will be repeatedly executed, the $^{13}$C longitudinal spin polarization is excited and allowed to evolve for a short $t_1$ increment (≈µs-ms, best performed in a constant-time 2D fashion for $t_1$-noise reduction) under the effects of a heteronuclear decoupling sequence that removes the effects of $^{1}$H$^{(proximate)}$ –while taking care to not significantly affect the bulk proton magnetization. At the end of the $t_1$ evolution time, the $^{13}$C magnetization is once again put into Hartmann-Hahn contact with its neighbouring proton, which had in the preceding $\tau_{mix}$ and $t_1$ periods been repolarized. During this CP process the $^{13}$C-coupled $^{1}$H$^{(proximate)}$ will be depolarized, by an amount that depends on the modulation imposed by the $t_1$ $^{13}$C chemical shift evolution (see Supporting Information, **Figures S1-S3**). Once again, care is here taken to ensure that the protons of the abundant spin pool, which are not dipolar-coupled to $^{13}$C, remain as unaffected by this CP contact as possible. (4) Following this point, the spin polarization of the $^{13}$C-coupled $^{1}$H$^{(proximate)}$, which has now been depleted by an amount that depends on the $^{13}$C's $t_1$ evolution, is repolarized *via* spin-diffusion over a new mixing time $\tau_{mix}$ in a manner analogous to that described above in step 2. However, a key difference in this case is that it will now be *the longitudinal spin polarization of the $^1H^{(abundant)}$ spin pool, that will be depleted by an amount depending on the $^{13}C$'s chemical shift $t_1$ evolution*. Repeating steps 3-4 numerous ($N$) times per scan up to a point dictated by the $T_1$ and $T_{1\rho}$ relaxation constraints of the $^1$H and $^{13}$C spin systems should thus lead to an appreciable depletion of the $^1$H polarization, in a $^{13}$C chemical-shift-dependent fashion. A final excitation pulse measuring the ensuing abundant $^1$H reservoir polarization then reveals the $^{13}$C $t_1$ modulation, which when incremented translates an entire $^{13}$C NMR time-domain (FID) signal as a modulation of the full proton reservoir response.

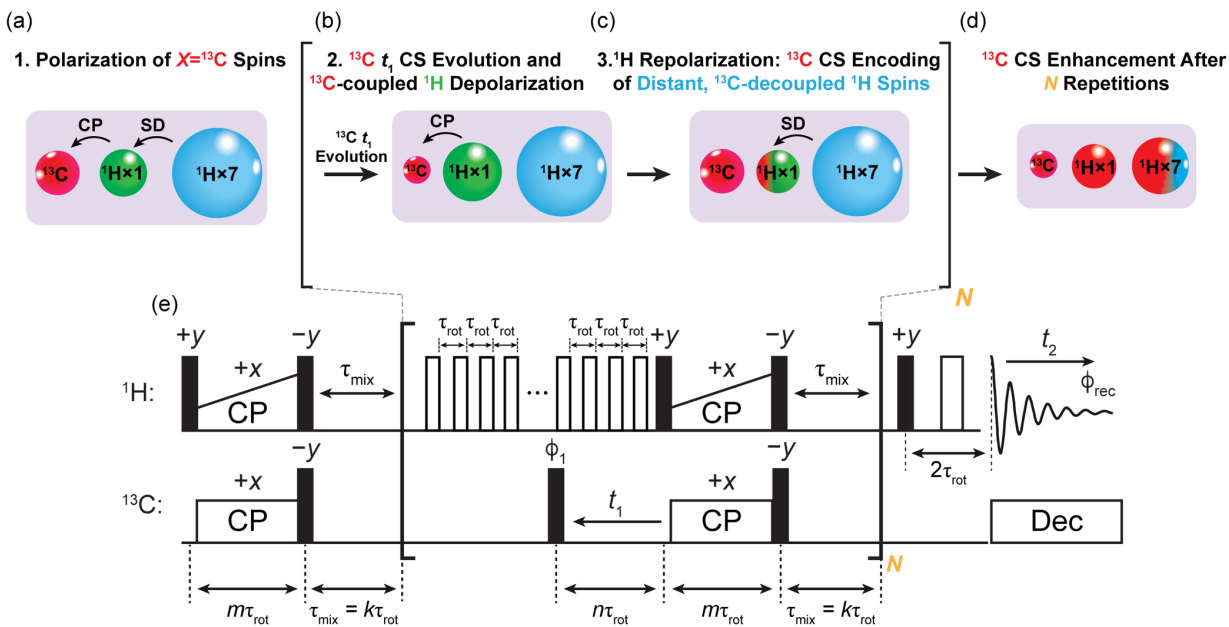

**Figure 1**. Schematic representation of the (a-d) spin dynamics over the course of the (e) Fourier-Encoded Saturation Transfer (FEST) pulse sequence. The magnetically dilute $^{13}$C spin pool, spatially-proximate $^{13}$C-coupled $^{1}$H spin pool, and the magnetically-abundant $^{13}$C-decoupled $^{1}$H spin pool, are represented in red, green, and blue, respectively. Colors are also used to represent the distinct spin pools and the size of the nuclei indicate the relative magnetizations. Cross polarization (CP) transfer and $^{1}$H-$^{1}$H spin diffusion (SD) processes are indicated with black arrows; see text for a more detailed explanation. Black and white pulses in (e) denote calibrated $\pi/2$ and $\pi$ pulses, respectively whose phases are indicated when relevant. The CP contacts along with the mixing interval, $t_1$ evolution time, and associated decoupling period are all rotor synchronized: $m$, $k$, and $n$ are integers and $\tau_{rot}$ is the rotor period. A rotor synchronized Hahn echo is also used to excite the final $^{1}$H observable.

*FEST NMR: Numerical Simulations*. This approach to amplify and detect $^{13}$C chemical shifts in a natural abundance solid was numerically tested with idealized quantum-based density matrix simulations,[38] carried out using custom-written code. These simulations were performed for a polycrystalline powder assumed to be undergoing MAS at $\nu_{rot}$ = 40 kHz, and for simplicity we assumed that the three spin pools schematically indicated by the three different colours in **Figure 1a-1d** possessed a distinct, uniquely addressable chemical shift: the abundant and dilute proton pools were set to $\nu(^1H)^{(abundant)}$ = 0 kHz and $\nu(^1H)^{(proximate)}$ = 10 kHz. These pools were composed of 7 and 1 $^1$H spins, respectively; a single $^{13}$C was dipole-coupled solely to the latter $^1$H and given an offset $\nu(^{13}C)$ = 2 kHz. The spin dynamics imposed by the FEST pulse sequence (**Figure 1e)** were then simulated on this spin ensemble and, *in lieu* of the final $^1$H excitation pulse, the expectation value of the abundant spin pool's $z$-polarization $\langle \hat{I}_z^{(abundant)} \rangle$ was monitored as a function of the number of FEST loops ($N$) and of the $^{13}$C $t_1$ evolution time. **Figure 2a** shows these expectation values, as calculated at the end of the last mixing interval in the sequence. The black curve in this figure shows the $t_1$-dependent trajectory of the $x$ $^{13}$C spin polarization component, $\langle \hat{S}_x^C(t_1) \rangle$, that results immediately after a single CP contact. This would be the conventionally detected signal in an indirectly-detected heteronuclear correlation experiment, and its clear 2 kHz frequency modulation serves as our reference. Calculated $\langle \hat{I}_z^{(\text{abundant})}(t_1) \rangle$ expectation values normalized with respect to this $\langle \hat{S}_x^C(t_1) \rangle$'s modulation as a function of looping are shown in other colours in this panel. Note that losses derived from pulse non-idealities and/or relaxation processes are here ignored, as are enhancements resulting from differences in the gyromagnetic ratios (*i.e.*, $\gamma(^1H)/\gamma(^{13}C)$). In other words, any modulation larger than ±0.5 represents a net signal enhancement over an indirect $^1$H-detected experiment. **Figure 2a** confirms that the $^{13}$C chemical shift modulation is ported with magnification onto the $^{13}$C-decoupled, abundant $^1$H spin pool; this corroborates **Scheme 1's** model, whereby a $^1$H→$^{13}$C→$^1$H CP transfer and subsequent $^1$H-$^1$H spin diffusion processes allow one to impart the $^{13}$C-modulation even on distant, dipolar-decoupled spin pools under fast MAS rotation. Notice that this $^{13}$C-derived, $^1$H$^{(\text{abundant})}$-detected modulation increases in amplitude as the number of loops *N* increases, as does the concomitant depletion of the abundant spin pool's polarization. At some point, however, the FEST looping leads to deviations from the ideal sinusoidal $t_1$-modulation; this $\langle \hat{I}_z^{(\text{abundant})}(t_1) \rangle$ distortion for large values of *N* reflects the small spin system here assumed and the lack of relaxation processes, which imposes non-linearities between the extent of the $^{13}$C-imposed modulation and the signal enhancement afforded by the $^1$Hs. While we have observed this behavior in certain solution-state, J-based analogues of this experiment (data not shown), we have not seen such distortions in the solid-state NMR measurements described below, presumably due to the extended nature of the spin-coupled solids network and the effects of spin relaxation.

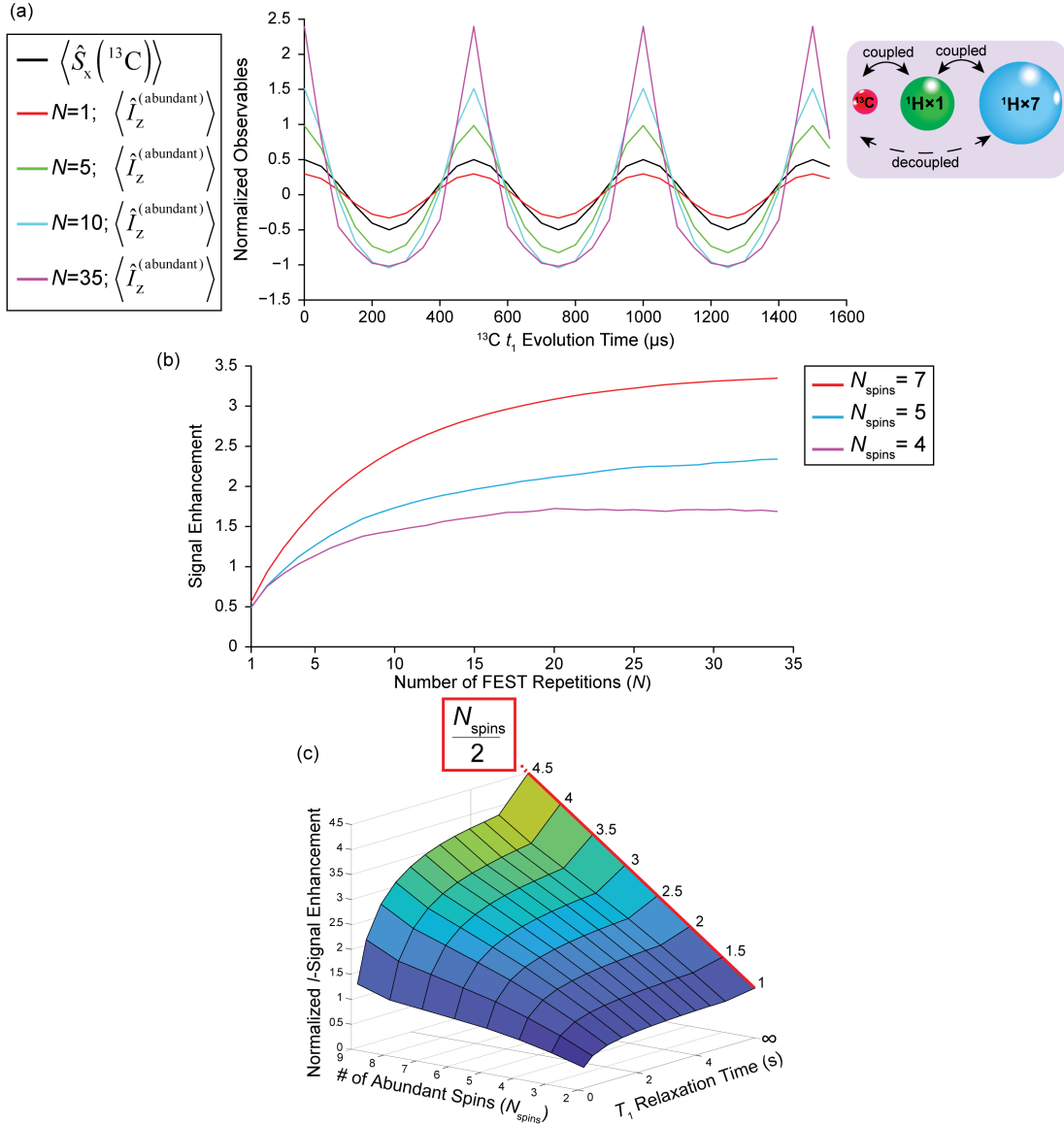

**Figure 2**. (a) Quantum mechanical (SIMPSON) simulations showing the time-dependent expectation values of $\langle \hat{S}_x^{(13C)} \rangle$ (black trace) for $^{13}$C after a single CP contact and $\langle \hat{I}_z^{(abundant)} \rangle$ for the seven protons comprising the abundant spin pool introduced in **Figure 1**. $\langle \hat{I}_z^{(abundant)} \rangle$ is calculated at the end of the mixing interval for different values of FEST repetitions ($N$). The isotropic shifts of the abundant and dilute spin pools were set to 0 kHz and 2 kHz, respectively and a $t_1$ increment of $\Delta t_1 = 50$ µs was used with a total of 32 $t_1$ points simulated. Homonuclear dipolar coupling constants of 2.5, 12, and 22 kHz were employed along with a single heteronuclear $^1$H-$^{13}$C dipolar coupling constant of 15 kHz. 100 kHz RF fields were used in all cases for all excitation and storage pulses and spin locking RF fields $\nu_1(^1\text{H}) = 100$ kHz and $\nu_1(^{13}\text{C}) = 60$ kHz were assumed for CP. Rotor synchronized radio-frequency driven recoupling (RFDR)[40] was used to achieve efficient homonuclear mixing in $\tau_{mix} = 4$ ms. Ideal decoupling was assumed during the $t_1$ evolution period by explicitly supressing the heteronuclear dipolar coupling Hamiltonian. A spinning speed of $\nu_{rot} = 40$ kHz was assumed and the average over the powder was calculated using 10 $(\alpha,\beta)$ crystallites sampled according to the REPULSION scheme.[41] (b) Signal enhancement of the $^{13}$C shift modulation detected in $\hat{I}_z^{(abundant)}$ as a function of $N$ and of the number of spins composing the abundant spin pool (*i.e.*, $N_{spins}$), arising from identical simulation parameters as in (a). (c) Normalized signal enhancements predicted for a two-site system incorporating a single modulated $I^{(proximate)}$ spin in chemical exchange with a larger $I^{(abundant)}$ reservoir containing $N_{spins}$, assuming four distinct exchange rates varying between 2-10 kHz. Bloch-McConnell equations were used to extract this as a function of $N_{spins}$ as well as of $T_1(^1\text{H})$ relaxation time (see the **Supporting Information** for further calculations and details).

**Figure 2b** examines the maximum enhancement of the $^{13}$C-modulation, showing that in the absence of relaxation it will ultimately depend on the number of spins in the abundant $^1$H pool. In general, the maximum modulation depth will be given by half the number of $^1$H spins in this pool (**Figure 2c, Supporting Figure S1**); given the $^{13}$C dilution at its natural abundance, this means that the potential $^{13}$C signal magnification afforded by FEST in organic solids can be very large. The actual conditions at which the maximum $^{13}$C enhancement will be achieved and its signal magnification value will depend on additional factors, including the $^1$H/$^{13}$C $T_1$ relaxation times and the effective internuclear dipolar couplings. Ancillary sets of simulations conducted on a simplified model system and pulse sequences, which both recapitulate and further explain these behaviors, are described in the Supporting Information (**Figures S1-S3).**

## Experimental Section

Samples of ibuprofen and sucrose were purchased from Sigma-Aldrich; L-Histidine HCl was purchased from B.D.H Laboratories. All were used as received without further purification. Samples were ground into fine powders and packed in 1.6 mm zirconium NMR rotors for measurement. NMR experiments were performed using a Varian VNMRS console interfaced to an Oxford 14.1 T ($v_0(^1H)$ = 600 MHz) wide-bore magnet. A Varian 1.6 mm triple-resonance HXY MAS probe was used for all NMR experiments. All spectra were acquired at a spinning speed of $v_{rot}$ = 40 kHz with a stability of ±5 Hz, using active pulse triggering and temperature regulation at 20° C. The magic angle of the probe was calibrated to 54.74° by maximizing the number of rotational echoes observed in the $^{81}$Br FID of KBr. $^1$H and $^{13}$C pulse width calibrations were performed using a sample of adamantane (40 kHz MAS), which was also used for chemical shift referencing. $^1$H→$^{13}$C{$^1$H} transfers were first calibrated using conventional CPMAS pulse sequences on each individual sample; contact times and spin-locking radiofrequency (RF) field strengths were then further refined in each 2D NMR experiment for the first $t_1$=0 increment. Rotary-resonance recoupling conditions employed in 2D indirectly-detected heteronuclear correlation (idHETCOR) experiments to suppress unwanted background signals were experimentally optimized by adjusting the length and power of the orthogonal recoupling pulses and then measuring the resulting $^1$H NMR spectra.[31] All 2D NMR experiments used the same heteronuclear π-based decoupling sequence, which gave optimized decoupling conditions, and had their $t_1$ evolution periods synchronized with the spinning frequency. Optimizing the FEST experiments required determining the mixing time (*i.e.*, $\tau_{mix}$) and number of loops (*i.e.*, $N$) combination that resulted in the largest overall depletion of the $^1$H spin polarization. This was done by repeatedly depolarizing proton magnetization *via* multiple-contact $^1$H-$^{13}$C CP for a fixed $N$/$\tau_{mix}$ combination, and then measuring the resulting $^1$H signal; this gave a so-called $S^{(on)}$ dataset. Repeating this under identical experimental conditions but in the absence of $^{13}$C spin-locking pulses gave a so-called $S^{(off)}$ dataset. Plotting {$S^{(off)}$-$S^{(on)}$}/$S^{(off)}$ as a function of both $N$ and $\tau_{mix}$ revealed the percentage of protons that was depleted; the largest value of this parameter gives the largest overall enhancement in terms of $^{13}$C SNR. Additional details pertaining to experimental optimizations are provided in the main text as well as in the Supporting Information (SI).

**Results**

*FEST NMR Experiments.* Having devised a way whereby $^{13}$C offsets can be imparted onto a distant abundant spin pool while under the effects of heteronuclear decoupling and MAS by exploiting the repeated depolarization/repolarization of a directly-bonded, mutually-coupled proton, a series of experimental tests were performed to corroborate these numerical predictions.

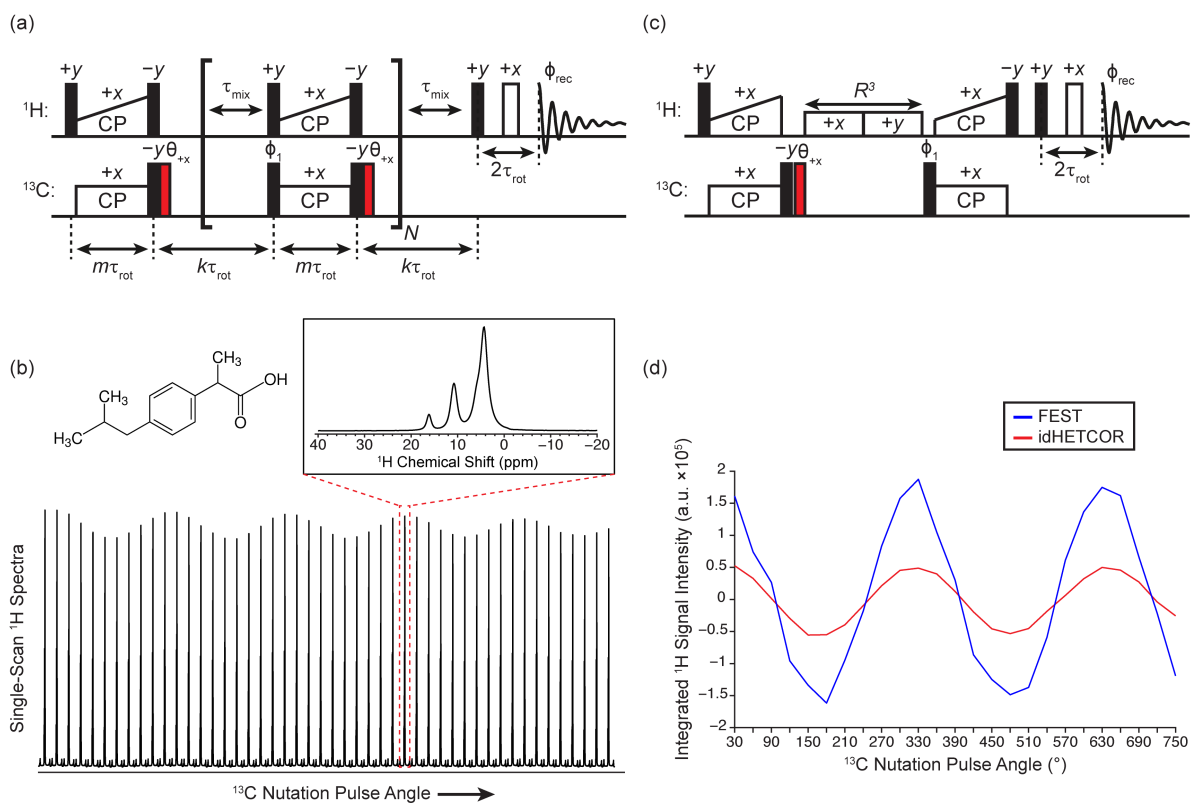

**Figure 3**. (a) Schematic representation of a modified FEST pulse sequence whereby a $^{13}$C nutation pulse (shown in red) encodes the $^{13}$C nutation frequency onto both $^{13}$C-coupled and $^{13}$C-decoupled protons. (b) Experimental $^1$H single-scan nutation-encoded FEST spectra collected using a naturally-abundant ibuprofen sample ($\nu_{rot}$ = 40 kHz, $B_o$=14.1 T) as a function of the $^{13}$C nutation angle, which is increasing in 30° steps from 0 to 1410°, for $N$ = 20 loops and $\tau_{mix}$ = 10 ms. The inset shows a representative $^1$H NMR spectrum at a particular $^{13}$C nutation angle. (c) Schematic representation of an optimized comparison consisting of a modified, nutation-encoded idHETCOR pulse sequence. This modified idHETCOR pulse sequence begins by exciting proton-coupled $^{13}$C spins with $^1$H-$^{13}$C CP and then encoding these same strongly-coupled protons with the $^{13}$C nutation frequency using $^{13}$C-$^1$H back CP. A rotary-resonance recoupling period placed between the CP contacts reintroduces proton homonuclear dipolar couplings that serve to rapidly de-phase any remaining transverse proton spin polarization, which ensures $^{13}$C polarization flows to proximate protons during back CP.[35] An added advantage of this dipolar recoupling period is an immunity against $t_1$ noise (*vide infra*). (d) Integrated $^1$H signal intensities collected with phase cycling (4 scans in both cases) using FEST (blue) and idHETCOR (red) on the same ibuprofen sample as (b).

These experiments were performed on the basis of auxiliary optimizations of the CP and mixing conditions, as provided in **Figure S2** of the Supporting Information. Shown as an initial experimental test is a simpler version of the FEST sequence (**Figure 3a**), where the $^{13}$C $t_1$ evolution is replaced by a variable-angle $^{13}$C nutation pulse θ (in red). This pulse modulates all $^{13}$C spins to the same extent (*i.e.*, at a single frequency), which can then be easily seen in the bulk, single-scan

¹H NMR signal as the ¹³C nutation pulse is incremented in a pseudo 2D fashion (**Figure 3b**). As expected the spin dynamics are such that when the ¹³C nutation angle is 0 or an integer multiple of $2\pi$, the ¹³C polarization that is spin locked in the looped CP processes is maximal; consequently, these angles result in the *least* amount of bulk ¹H depolarization. By contrast, when $\theta(^{13}C)$ is an integer multiple of $\pi$, the ¹³C spin polarization is stored along the $-z$ axis after the first CP and is therefore, antiparallel with respect to its spin locking $B_1$ field during subsequent Hartmann-Hahn contacts; this results in the *largest* depolarization of the ¹H bulk signal (see **Figure S3** and the discussion therein for further details). Notice that all other values of $\theta(^{13}C)$ depolarize the bulk ¹H spins in the expected $\cos(\theta)$-dependent manner, with no evidence of the small-reservoir distortions noted in **Figure 2**; this is a consequence of the small ¹³C/¹H ratio in this polycrystalline natural-abundance sample. Notice as well the strength of the FEST effect, which is sufficient to easily detect the ¹³C nutation frequency in these single-scan bulk proton signals, and is akin to what we have observed in water-based ¹H CEST observations of non-labile ¹³C NMR spectra.[28] It is possible to approximately estimate the degree of the enhancement in the FEST MAS experiments, by comparing their modulation against that which can be indirectly detected in ¹³C-filtered polarization, by appropriate phase cycling of the ¹³C excitation pulse that precedes CP. The result of this is a modified idHETCOR pulse sequence,[34] whereby the $t_1$ evolution period has been replaced with a ¹³C nutation pulse (**Figure 3c**). **Figure 3d** shows *ca.* a 3× signal enhancement achieved with FEST over an idHETCOR, with both datasets acquired under similarly optimized experimental conditions.

Experimental tests were also performed to explore FEST's ability to read-out site-resolved ¹³C NMR spectra in this manner. **Figure 4** compares FEST NMR results obtained on a polycrystalline sample of naturally-abundant sucrose, *vs.* results from an optimized version of the idHETCOR sequence (**Figure S4**). Additional comparisons collected on naturally abundant ibuprofen and L-histidine HCl samples, are summarized in the Supporting Information (**Figures S5-S6**). The resulting 2D FEST spectrum (**Figure 4a**) consists of the single broad peak characteristic of the bulk $^1$Hs observed under MAS conditions along the direct ($F_2$) dimension, showing strong correlations to seven $^{13}$C resonances observed in the indirect ($F_1$) dimension. A signal amplification of *ca*. 7× is achieved against the optimized 2D idHETCOR counterpart (**Figure 4b**), which is also visible by comparing the $^1$H NMR spectra acquired for the first $t_1$=0 increment of FEST and idHETCOR (**Figure 4c**, top and bottom, respectively). More modest (3-5×) FEST enhancements were observed on the ibuprofen and histidine HCl samples, probably because of the latter's shorter $T_1(^1H)$ values (**Figures S5-S6**). Notice that in all cases, the FEST experiment highlights both protonated and non-protonated $^{13}C$s owing to its reliance on relatively long CP times.

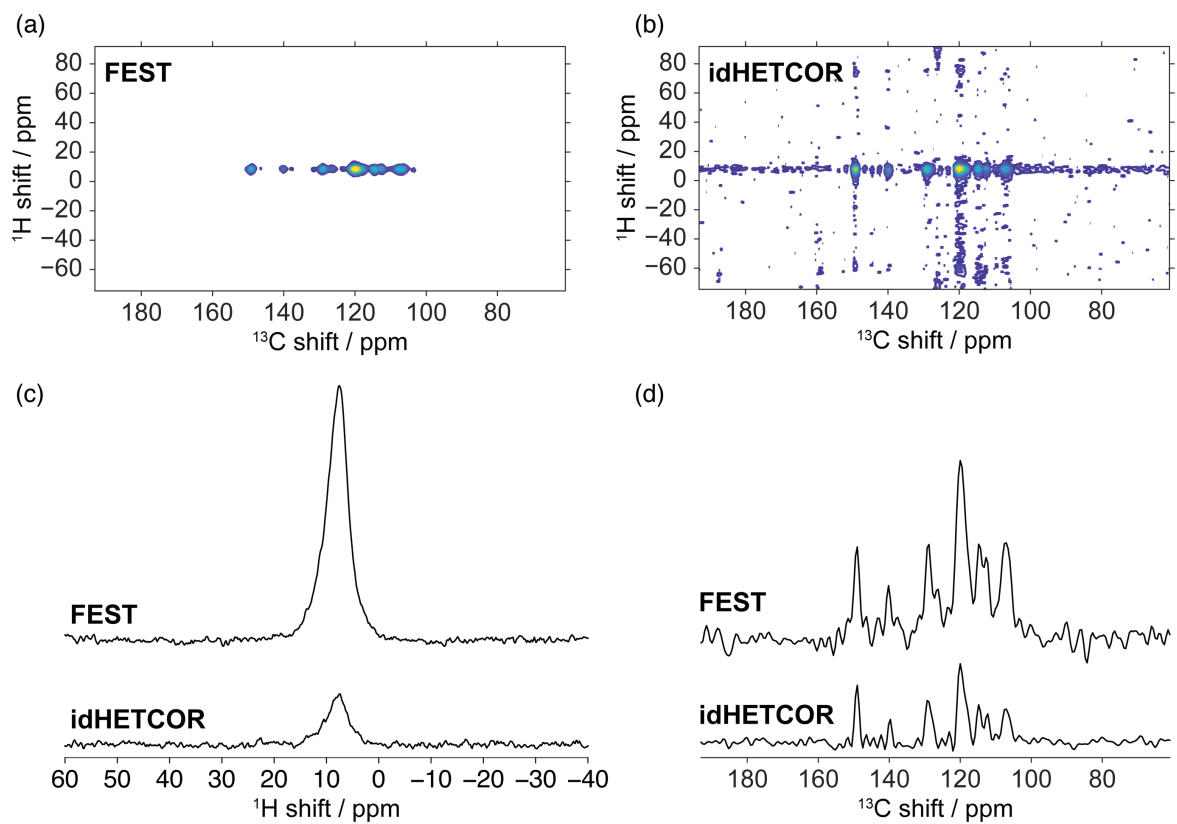

**Figure 4**. Experimental proton-detected $^1$H-$^{13}$C correlation spectra of naturally abundant sucrose ($v_{rot}$ = 40 kHz) collected with (a) FEST ($\tau_{mix}$ = *ca*. 100 ms, *N*=15) and (b) idHETCOR in 2 scans and 4 scans, respectively (recycle time is 60 s for both and a total of 64 $t_1$ points were collected under optimized experimental conditions). (c) $^1$H NMR spectra collected for the first $t_1$ increment using (top) FEST and (bottom) idHETCOR. (d) $^{13}$C $F_1$ slices taken from the same $F_2$ point in the 2D spectra for (top) FEST and (bottom) idHETCOR. The minimum, maximum, and spacing of contour levels is set based on the standard deviation of the noise in each dataset.

The primary challenge facing the sensitivity enhancement achievable from these $^{13}C$-shift-encoded FEST experiments, turned out to be choosing an optimized heteronuclear dipolar decoupling sequence during the $t_1$ evolution period. Such a sequence has to deliver a high-resolution $^{13}C$ NMR spectrum without $^1H$ coupling artifacts or broadenings, while at the same time preserving and leaving the abundant $^1H$ spin polarization "untouched", which will eventually be the source of any $^{13}C$ FEST enhancement. Any scan-to-scan instability in the performance of this heteronuclear $t_1$-decoupling sequence, led to unacceptable amplifications in the $t_1$ noise. Our final selection was a π-based, constant-time heteronuclear dipolar decoupling sequence, involving repeated units of 8 short, high-power, rotor-synchronized $^1H$ π-pulses applied at the end of a rotor period, with their phases XY8 phase-cycled.[37,39] The overall number of $^1H$ decoupling pulses were thus kept constant, and the effective $^{13}C$ evolution time was defined by "walking" the $^{13}C$ excitation pulse throughout this period, in increments matching the XY8 decoupling subunits. This procedure led to penalties both in terms of sensitivity losses associated to the constant-time operation, partial saturation of the $^1H$ reservoir, and the introduction of scan-to-scan instabilities leading to $t_1$ noise –but it was the best we managed to devise.

## Discussion and Conclusions

The present study put forward a time-domain, CEST-inspired experiment, for increasing the sensitivity of solid-state HETCOR NMR experiments. Unlike liquid-state CEST, which relies on actual chemical exchanges to facilitate the transfer of chemical shift information between a magnetically dilute spin pool and an abundant water spin pool, the FEST experiments here presented rely on an abundant $^1$H reservoir for achieving its sensitivity enhancement, and on the discriminated use of heteronuclear and homonuclear dipolar couplings for driving the various transfers of polarization/saturation. The former is used, *via* CP contacts, to translate the heteronuclear chemical shift encoding into proton depolarization, while spin diffusion is used to extend this depletion throughout distant protons, for the sake of achieving sensitivity enhancement. In this fashion, FEST experiments performed on naturally-abundant organic compounds allow for the nutation or shift modulation of a magnetically dilute $^{13}$C spin pool to be encoded onto a magnetically abundant $^1$H spin pool. In parallel to what was exploited in the solution-state REFLEX experiment,[28] a defining feature of FEST is the conceptual separation between magnetically dilute $^1$Hs that are strongly dipolar-coupled with the heteronucleus, and the abundant, essentially $^{13}$C-decoupled $^1$H spin reservoir whose depletion will eventually lead to the signal enhancement. In the solution-state NMR case, the former were given by labile $^1$Hs that are *J*-coupled to the heteronucleus, but also capable of exchanging with the water pool; in the latter their definition is less clear, as the "dilute" $^1$Hs are defined by the pool of spins that can cross-polarize to/from a given $^{13}$C site, while the "abundant" $^1$Hs are those parts of the larger network that has so far remained mostly passive in heteronuclear NMR. In the case of naturally-abundant organic solids it is clear that the latter will entail a larger majority of protons, opening a new route to a more efficient use of these abundant species' polarization for enhancing heteronuclear NMR. The present study relied on spontaneous spin-diffusion for the relay of polarization between the dilute and abundant $^1$H spin reservoirs; it is conceivable that, particularly at the faster MAS rates that should facilitate this experiment, RF-assisted approaches might enhance the effectiveness of this process (in fact the train of π-pulses inserted during the $t_1$ period for heteronuclear decoupling, facilitated this *via* an RFDR-like mechanism).

As mentioned, the main limitation of the sequence that we devised was its high sensitivity towards $t_1$ noise artifacts, which clearly lowered the apparent gains of the experiment. These arose from fluctuations in the scan-to-scan intensities of the bulk $^1$H NMR signals, which we ascribe mostly to inconsistencies brought about by the hundreds of $^1$H π-pulses applied to achieve heteronuclear decoupling during $t_1$. These instabilities greatly decreased upon rotor-synchronizing the whole sequence (**Figure S7**), which included the synchronization of all events to the same initial sample rotor phase and by actively regulating the sample temperature (**Figure S8**). Nevertheless, despite these efforts, multiplicative $t_1$ noise artifacts were still observed in the final bulk $^1$H signal. This can be clearly appreciated in the $^{13}$C $F_1$ trace taken from the 2D FEST spectrum (**Figure 4d**, top) that, unlike the idHETCOR counterpart (**Figure 4d**, bottom), features strong $t_1$ noise ridges. Furthermore, the application of common strategies like rotary-resonance recoupling or phase cycling procedures to destroy $^1$H spin polarization that has not been modulated over $t_1$, cannot be used in FEST, as this would eliminate any potential signal enhancement. Alternative methods to attenuate the $F_1$ artifacts are currently being developed and tested. These include saturation-transfer methods that progressively deplete the abundant proton reservoir while invoking dipolar-order-mediated CP –a strategy proven effective at reducing $t_1$-noises when targeting NMR spectra under static conditions.[43] Such dipolar-order states have also been reported for rotating solids,[44] opening an interesting alternative for FEST experiments. Using such protocols under conditions of fast ($v_{rot}$ > 60 kHz) MAS could also alleviate the need for heteronuclear decoupling during the $t_1$ evolution, thereby greatly reducing the $t_1$ noise distortions affecting the $^1$H-decoupled CP FEST experiments here described. Alternative sequences performed at fast spinning rates but operating on the basis of *J*- instead of dipolar-transfers, could further reduce $t_1$ noise artifacts. Efforts are underway to implement these, as well as customized denoising approaches based on data post-processing routines. Despite these current limitations and needs for improvement, FEST seems to open hitherto untapped sensitivity enhancements in solid-state high-resolution NMR, by the efficient use of abundant spin polarization that typically goes unused. Extensions of similar ideas to solution-phase experiments involving natural-abundance (non-exchanging) organic substrates, are also being explored.

**Supporting Information**. Numerical simulations of FEST MAS NMR on a model solid-state homonuclear system; additional experimental details on setting up the FEST MAS NMR experiment; additional experimental examples

**Notes.** The authors declare no competing financial interests.

**ACKNOWLEDGMENT.** We are grateful to the late Koby Zibzener for technical assistance in the experiments here described; we also acknowledge fruitful discussions with Dr. Rob Tycko (NIDDK, NIH) regarding this work. This research was supported by the Israel Science Foundation (grant #965/18), the EU Horizon 2020 program (Marie Skłodowska-Curie Grant 642773 and FET-OPEN Grant 828946, PATHOS), and the generosity of the Perlman Family Foundation. LF holds the Bertha and Isadore Gudelsky Professorial Chair and Heads the Clore Institute for High-Field Magnetic Resonance Imaging and Spectroscopy, whose support is acknowledged.

**Supporting Information for**

# "On the Potential of Fourier-Encoded Saturation Transfers for Sensitizing Solid-State Magic-Angle Spinning NMR Experiments"

Michael J. Jaroszewicz, Mihajlo Novakovic, and Lucio Frydman*

*Department of Chemical and Biological Physics, Weizmann Institute of Science,*

*Rehovot, Israel, 7610001*

* E-mail: lucio.frydman@weizmann.ac.il

### 1. FEST MAS NMR: Numerical simulations on a model solid-state system

Prior to assessing the performance of heteronuclear FEST, simulations were carried out on the model $^1$H system shown in **Figure S1a**. Inspired by analogous liquid-state CEST experiments, these dilute (red) and abundant (blue) spin pools are composed of 1 and 7 protons, respectively. The chemical shift of the dilute spin pool was set to $v_{iso}$ = 6 kHz and the abundant spin pool was set to $v_{iso}$ = 0 kHz. These two spin pools were assumed dipolar coupled by interactions as those described in **Figure 1**. Shown as well in **Figure S1b** is the NMR spectrum expected to arise under these conditions upon performing MAS at 40 kHz; also shown is a preliminary experiment seeking to analyze potential sensitivity enhancements in FEST. The setup and pulse sequence of this experiment (**Figure S1c**) is that of the liquid-state FLEX NMR experiment;[1,2] however, a key difference is that the chemical shift modulation of the dilute $^1$H pool will be transferred to the abundant spin pool *via* dipolar-driven spin diffusion rather than by chemical exchange. The magenta curve in **Figure S1d** oscillating between ±1, is the conventional FID expected from the dilute spin pool as measured immediately after the $t_1$ evolution period: its 6 kHz modulation and its depth serve as the standard for calculating potential enhancements. The green curve centered at 7 along the y-axis shows the $t_1$-evolved expectation of the abundant spin pool $\langle \hat{I}_z^{(abundant)} \rangle$ after N=1 loop, showing that in the absence of homonuclear dipolar couplings between the two spin pools there is no transfer of chemical shift information (the isotropic shift of the abundant spin pool having been set to 0). By contrast, simulations under identical conditions reveal that when the homonuclear dipolar coupling Hamiltonian is active (**Figure S1d**, red curve) a clear 6 kHz modulation appears on $\langle \hat{I}_z^{(\text{abundant})} \rangle$ even for $N$=1. This modulation increases upon increasing the number of loops $N$ (blue, purple, black lines), up to the point when the modulation equals the number of abundant spins / 2, which is the maximum that can be achieved for a small spin system. This is more clearly illustrated by the plots shown in **Figure S1e**, which shows the modulation depth of $\langle \hat{I}_z^{(\text{abundant})} \rangle$ as a function of the number of loops for different sizes of the abundant spin pool. It is also enlightening to evaluate how this model predicts the repolarization of the dilute $^1$H pool to proceed over the course of the mixing periods; **Figure S2** demonstrates this repolarization, at rates and extents that naturally will depend on the number of loops. Notice that in general, using a small number of loops will lead to depolarizations of the bulk reservoir that are linear with the depth of the originating modulation (red curve, **Figure 2**); as $N$ is increased however, so as to maximize the depletion, the depletion becomes non-linear, which results in the non-sinusoidal distortions observed for large $N$ in **Figures 2** and **S1d**. These distortions were not visible in the nutation experiments performed in regular organic solids (*e.g.*, **Figure 3**), as they arise solely in small spin systems and the absence of relaxation. In summary, these homonuclear FEST simulations show the possibility of encoding and detecting the offset modulations of a dilute spin pool in a more highly populated abundant spin pool for a rotating polycrystalline powder. The transfer of polarization is facilitated by spin diffusion and leads to a depletion of the abundant spin pool's polarization in a magnified, $t_1$-dependent fashion, that unlike the CW CEST experiment is both Fourier encoded and does not require exact knowledge of the dilute spin's resonance frequency. The maximum enhancements will be dictated by the total number of spins comprising the abundant spin pool and the number of repeated loops; eventually, in finite-sized simulations, this may reach a non-sinusoidal behavior as the kinetics of the depolarization/repolarization processes saturate and become non-linear.

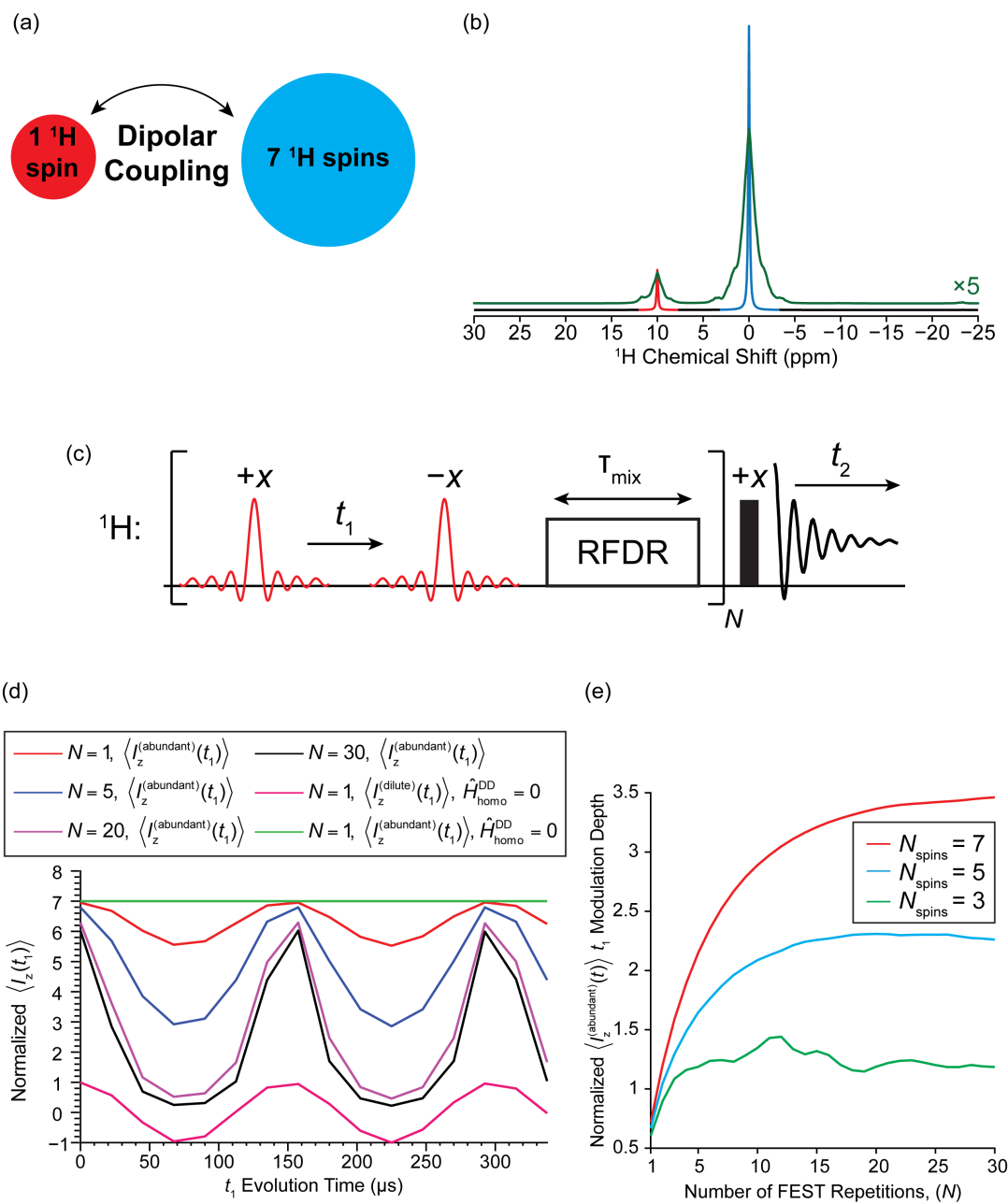

**Figure S1**. (a) Schematic representation of how an organic solid can be formally divided into two unequally populated and individually addressable spin pools whose resulting MAS NMR spectra are shown in (b) when the homonuclear couplings are suppressed (black line) and active (green line). (c) Schematic representation of the homonuclear $^1$H FEST pulse sequence, which features a looped $t_1$-incremented block that selectively acts on the dilute spin pool (schematically indicated by the red sinc pulses) followed by a homonuclear mixing interval. After $N$ loops, the entire $^1$H spin system is excited, seeking to reveal an enhanced chemical shift modulation of the dilute spin pool detected on the abundant $^1$H spin pool resonance. All simulations derived from this sequence assumed 40 kHz MAS and were carried out using SIMPSON[3]. (d) SIMPSON numerical simulations of the time-dependent $t_1$ expectation values of the z-component of spin polarization for the dilute (denoted $\langle \hat{I}_z^{(\text{dilute})} \rangle$) and abundant (denoted $\langle \hat{I}_z^{(\text{abundant})} \rangle$) $^1$H spin pool observables over the course of the FEST pulse sequence.

**Figure S1 (CONT.)** Virtually the same simulation parameters are used here as those used in the main text, with the exception that the heteronuclear $^{13}$C spin was removed. The initial density matrix was set to $\hat{\rho}(t=0) = \hat{I}_z^{(\text{dilute})} + \hat{I}_z^{(\text{abundant})}$ (*i.e.*, a fully polarized $^1$H spin system) and the detection state was again selective to observe only the abundant spin pool, unless stated otherwise. The expectation values are normalized with respect to the number of spins comprising each respective spin pool and are measured at the end of the mixing interval, unless stated otherwise. (e) FEST signal enhancements calculated as a function of the number of loops for the abundant spin pool for three different populations.

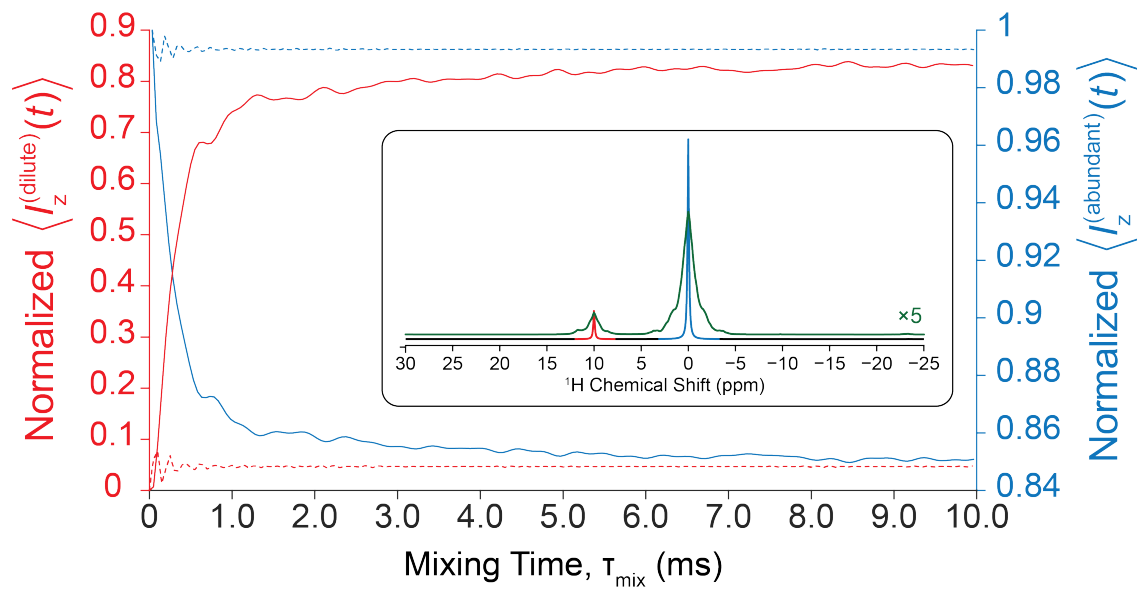

**Figure S2.** SIMPSON numerical simulations showing the time-dependent repolarization and depolarization of the dilute (red) and abundant (blue) spin pools, as a result of longitudinal dipolar-driven proton spin diffusion. The solid lines were simulated using radio-frequency driven recoupling (RFDR, 8 repeating 200 kHz π pulses applied every rotor period with phases *x*, *y*, *y*, *x*, −*x*, −*y*, −*y*, −*x*) whereas the dashed lines were simulated in the absence of RFDR. In this case, RFDR provides the most efficient spin diffusion, which is required due to the fact that the abundant and dilute spin pool resonances are well resolved from each other, even when the full homonuclear dipolar coupling Hamiltonian is active (the corresponding $^1$H NMR spectra that result when the homonuclear dipolar interaction is active and suppressed are shown in the inset with colours green and black, respectively). All curves were calculated by detecting the $\langle \hat{I}_z \rangle$ expectation value of the respective spin pools in which the initial density matrix was set to a fully polarized abundant spin pool and a completely depleted dilute spin pool (*i.e.*, $\hat{\rho}(t=0) = \hat{I}_z^{(\text{abundant})}$), which mimics fully efficient CP transfer (*i.e.*, the instance when the $^{13}$C-coupled proton transfers all of its spin polarization to $^{13}$C and therefore its value of $\hat{I}_z$ is effectively 0 before homonuclear mixing). After about 4 ms of homonuclear mixing, roughly 80 % of the dilute spin pool's polarization is recovered at the expense of a corresponding *ca*. 14 % drop of the abundant spin pool's polarization. This process of depleting the abundant spin pool *via* dipolar-driven repolarization of an amplitude-modulated $^1$H spin pool is the primary mechanism underlying FEST.

*Including the $^{13}C$: Polarization transfers on a model three-site solid-state system.* **Figure S3** follows the key polarization transfer processes that happen in a heteronuclear FEST experiment involving three distinct pools: the $^{13}C$ dilute spin pool ($\langle \hat{S}_z \rangle$, denoted with a diamond), the $^{13}C$-coupled $^1H$ spin pool ($\langle \hat{I}_z \rangle$, denoted with a circle), and a "$^{13}C$-decoupled" abundant $^1H$ spin pool ($\langle \hat{F}_z \rangle$, denoted with a triangle). Within every CP period $^1H$ spin polarization flows from the $^{13}C$-coupled proton (circle) to the $^{13}C$ spin (diamond), which in turn causes a depletion of the abundant spin pool's polarization after every subsequent mixing interval. For instance, starting at the first observation point, which is depicted in blue, $\langle \hat{S}_z \rangle$ for $^{13}C$ (diamond) is larger than $\langle \hat{I}_z \rangle$ of the $^{13}C$-coupled proton (circle) resulting from the first CP transfer; however the former is only slightly larger than the latter because the CP transfer is not fully efficient resulting in *ca.* 60% $^{13}C$ polarization. The abundant $^1H$ spin pool (triangle) immediately after the first CP contact (after *z*-storage) remains fully polarized at a value of 7 (corresponding to 100% polarization), as this spin pool is not heteronuclear dipolar coupled to $^{13}C$ and therefore does not participate in the CP. After homonuclear mixing ($\tau_{mix}$ = 4 ms), proton spin diffusion causes the spontaneous repolarization of the $^{13}C$-coupled proton $\langle \hat{I}_z \rangle$ from *ca.* 40 to *ca.* 90 % (going from the blue shapes to the adjacent green shapes) along with a concomitant depletion of the abundant spin pool's longitudinal polarization $\langle \hat{F}_z \rangle$ from 100 to *ca.* 92%. Meanwhile the $^{13}C$ spin polarization $\langle \hat{S}_z \rangle$ remains constant, as it does not undergo any evolution whilst the effects of relaxation are ignored. Comparing the expectation values going from the green time point to the red time point (*i.e.*, to the time point after $t_1$ evolution), it is clear that neither the abundant $^1H$ nor the $^{13}C$-coupled $^1H$ spin pool undergo any evolution. The $^{13}C$, however, undergoes a chemical shift evolution, leading to a drop in $\langle \hat{S}_x \rangle$ of *ca.* −60%, resulting from the $t_1$= 225 μs evolution time (corresponding to an absolute local minimum for the 2 kHz $^{13}C$ chemical shift assumed). During the following CP contact, the $^{13}C$ spin-locking $B_1$ field, which is applied along the +*x* axis and is largely anti-parallel with respect to the $^{13}C$ transverse spin polarization, results in the largest depletion of $^{13}C$-coupled $^1H$ spin polarization; a concomitant change in the $^{13}C$ $\langle \hat{S}_x \rangle$ value from *ca.* −60% to *ca.* 55% polarization is observed. Finally, when looking at the change in $\langle \hat{F}_z \rangle$ for the abundant $^1H$ spin pool, a roughly 6% depletion of its overall value has occurred due to the repolarization of the $^{13}C$-coupled proton. These spin dynamics continue for the following loops, confirming the assumptions stated in the main text concerning the CP-dependent depolarization and subsequent amplitude modulation of the $^{13}C$-coupled proton.

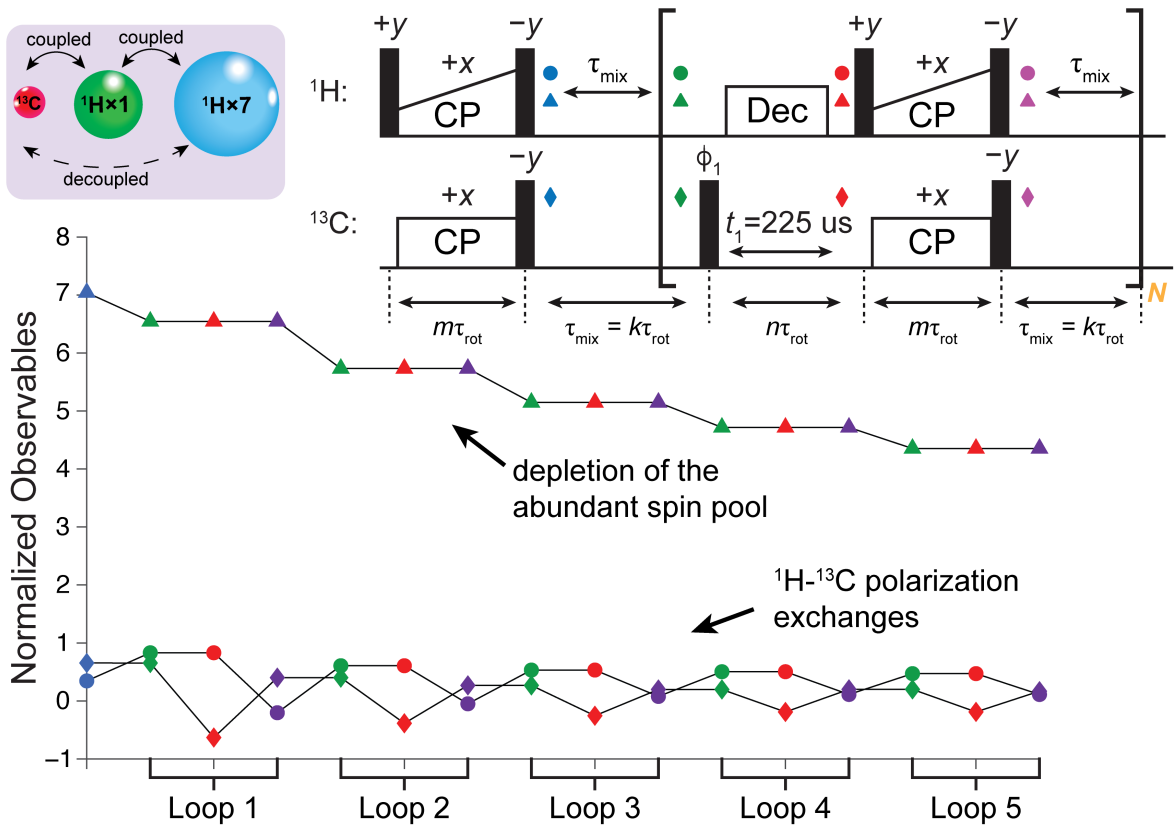

**Figure S3**. Expectation values for each of the three spin pools (shown as an inset) measured over the course of the FEST pulse sequence (shown as an inset) for the first 5 consecutive loops (indicated on the *x*-axis). The inset also indicates the time points at which each of the expectation values are evaluated with the use of the colour-coded geometrical shapes. For each loop and for each of the 3 expectation values, 4 data points are acquired at different time points, giving a total of 12 calculated expectation values. The 4 distinct colours refer to the 4 different time points at which a particular expectation value is evaluated in a given loop, and the different geometrical shapes refer to the 3 different expectation values that are calculated, which are $\hat{I}_x$ of the $^{13}$C dilute spin pool (*i.e.*, $\langle \hat{I}_x \rangle$ and denoted with a diamond), $\hat{I}_z$ of the $^{13}$C-dipolar coupled $^1$H spin pool (*i.e.*, $\langle \hat{I}_z \rangle$ and denoted with a circle), and $\hat{I}_z$ of the $^{13}$C-decoupled abundant $^1$H spin pool (*i.e.*, $\langle \hat{F}_z \rangle$ and denoted with a triangle). In this case, the expectation values are normalized with respect to the number of spins composing each of the respective spin pools. For instance, $\langle \hat{I}_x \rangle$ and $\langle \hat{I}_z \rangle$ for the dilute and $^{13}$C dipolar-coupled $^1$H spin pool, respectively reach maximum and minimum values of *ca.* ± 1, whereas $\langle \hat{I}_z \rangle$ for the abundant spin pool takes on a maximum value of 7. The same simulation parameters are employed in this figure as those used in **Figure 1** in the main text, with the only exception being that only a single $t_1$ time is used.

## 2. FEST NMR: Additional experimental details and examples

The subsequent three figures present further details on the pulse sequences employed in this study (**Figure S4**) and two experimental datasets collected on the naturally-abundant organic solids ibuprofen and L-Histidine HCl. These experimental datasets compare and contrast the performance of FEST with that of the optimized idHETCOR pulse sequence in terms of experimental time, SNR, as well as the overall spectral quality.

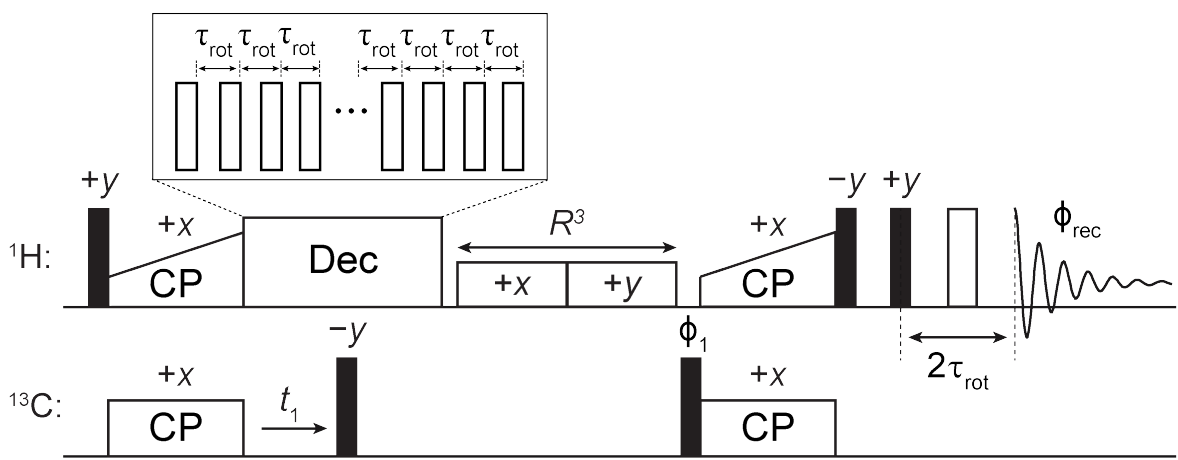

**Figure S4**. Schematic representation of the indirectly-detected HETCOR (idHETCOR) pulse sequence that is used as an optimized comparison against 2D FEST spectra. An initial proton $\pi/2$ pulse excites transverse spin polarization that is then transferred to dipolar-coupled and proximate $^{13}$C nuclei. After a sufficiently long contact time, the polarized $^{13}$C spins then undergo chemical shift evolution for an incremented $t_1$ time under the effects of the same heteronuclear $\pi$-based decoupling scheme used in the FEST experiments (*i.e.*, in a constant-time fashion). The $t_1$-encoded $^{13}$C spin polarization is then stored along the $+z$ axis and a period of rotary-resonance recoupling ($R^3$) is used to de-phase all remaining transverse and longitudinal $^1$H spin polarization *via* the reintroduction of homonuclear dipolar couplings. $R^3$ is performed by applying two long and low-power CW pulses with orthogonal phases whose combined timing is an integer multiple of the rotor period and whose RF field is equal to half the spinning frequency (*i.e.*, $\nu_1 = 0.5\nu_{rot}$). A $^{13}$C excitation pulse followed by back $^{13}$C-$^1$H CP contact transfers the $^{13}$C CS-encoded spin polarization to the infinitely hot proton system where it is ultimately detected with a rotor-synchronized Hahn echo.

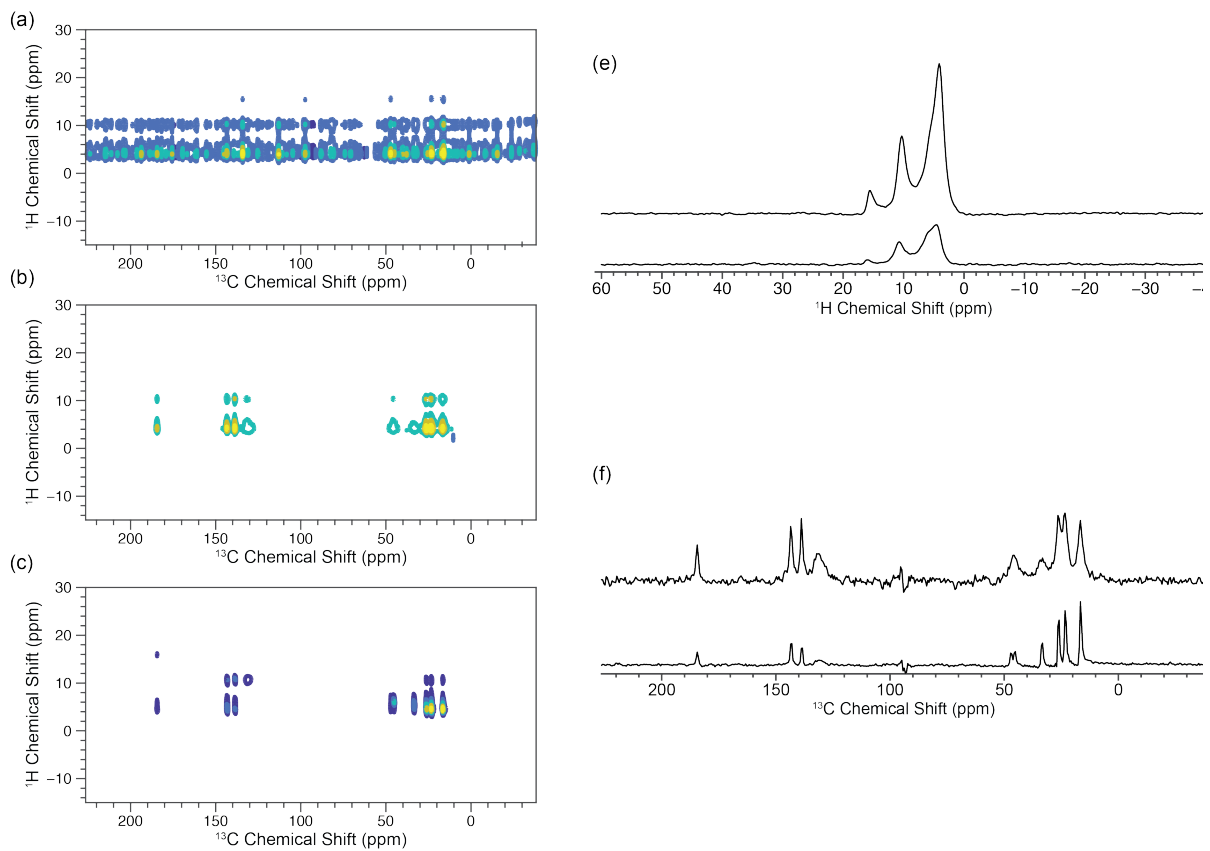

**Figure S5**. Experimental proton-detected $^1$H-$^{13}$C correlation spectra of naturally abundant ibuprofen ($\nu_{rot}$ = 40 kHz) collected with (a) FEST (with $F_1$ decoupling), FEST (without $F_1$ decoupling), and (c) idHETCOR (with $F_1$ decoupling) all in 4 scans, respectively (recycle time is 2 s in all cases and a total of 256 $t_1$ points were collected under optimized experimental conditions). (e) $^1$H NMR spectra collected for the first $t_1$ increment using (top) FEST (without $F_1$ decoupling) and (bottom) idHETCOR (with $F_1$ decoupling). (f) $^{13}$C $F1$ slices taken from the same $F_2$ point in the 2D spectra for (top) FEST (without $F_1$ decoupling) and (bottom) idHETCOR (with $F_1$ decoupling). Comparing the $F_2$ $^1$H spectra reveals a signal enhancement of *ca.* 3.5×. The FEST spectrum acquired with $F_1$ decoupling shows the artifacts that arise upon π-based decoupling.

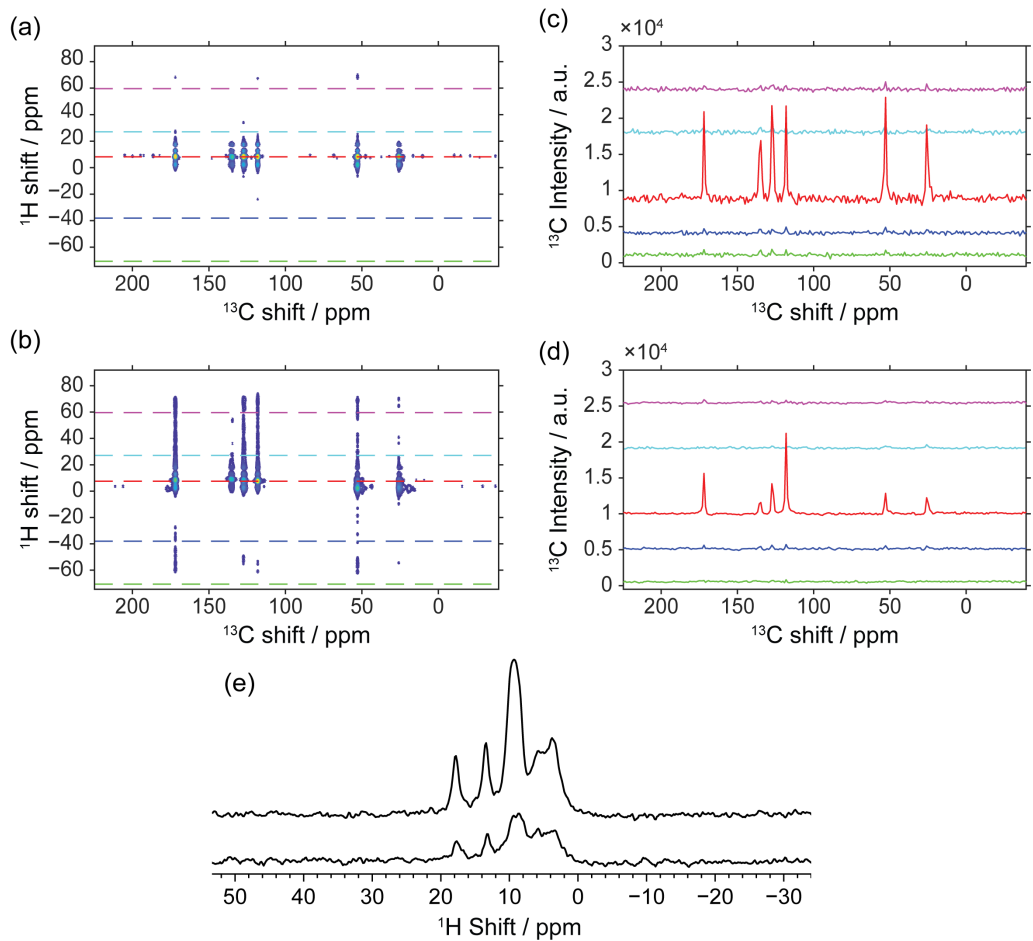

**Figure S6**. Experimental proton-detected $^1$H-$^{13}$C correlation spectra of naturally abundant L Histidine HCl ($\nu_{rot}$ = 40 kHz) collected with (a) FEST and (b) idHETCOR all in 4 scans with $F_1$ decoupling, respectively (recycle time is 3 s in all cases and a total of 256 $t_1$ points were collected under optimized experimental conditions). Various $^{13}$C $F1$ slices (indicated by the horizontal dashed lines) taken from the same $F_2$ points in the (c) FEST and (d) idHETCOR spectra. $^1$H NMR spectra collected for the first $t_1$ increment using (top) FEST and (bottom) idHETCOR. Comparing the $F_2$ $^1$H spectra reveals a sensitivity enhancement of *ca*. 3× and a more uniform enhancement of the $^{13}$C FEST peaks, albeit with larger $t_1$ noise artifacts.

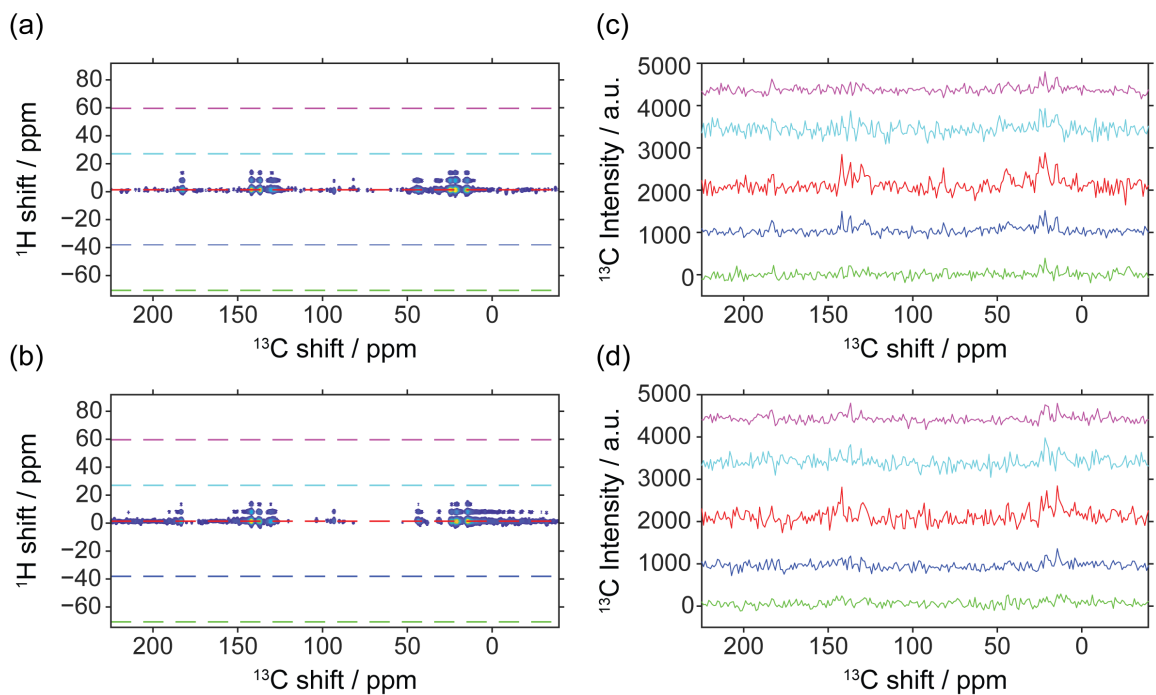

**Figure S7**. Experimental (left) 2D FEST spectra of ibuprofen and the corresponding $^{13}C$ $F_1$ slices recorded in the (a, c) presence and absence (b, d) of active MAS pulse triggering.

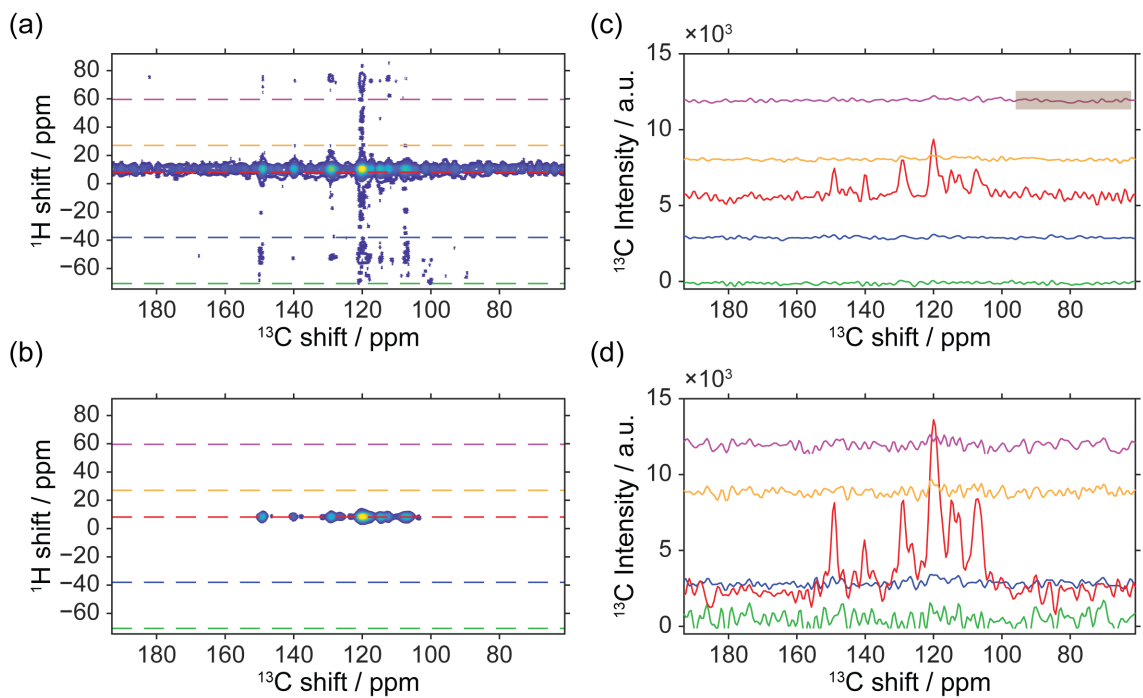

**Figure S8**. Experimental (left) 2D FEST spectra of sucrose and the corresponding $^{13}$C $F_1$ slices recorded in the (a, c) absence and presence (b, d) of active temperature regulation.